\documentclass[twocolumn,showpacs,amsmath,amssymb]{revtex4}

\usepackage{latexsym}
\usepackage{amssymb}
\usepackage{amsfonts}
\usepackage{amsmath}
\usepackage[dvips]{graphicx}
\usepackage{times}
\usepackage{colordvi}

\begin{document}
\bibliographystyle{unsrt}

\title{On the iterated Crank-Nicolson for hyperbolic and parabolic
  equations in numerical relativity}

\author{Gregor Leiler$^{1,2}$, and Luciano Rezzolla$^{2,3,4}$}
      
\affiliation{$^{1}$ Department of Physics, Udine University, Udine, Italy}

\affiliation{$^{2}$Max--Planck--Institut f\"ur Gravitationsphysik, Albert
Einstein Institut, Golm, Germany}

\affiliation{$^{3}$ SISSA, International School for Advanced Studies and
INFN, Trieste, Italy}

\affiliation{$^{4}$ Department of Physics, Louisiana State University,
Baton Rouge, USA }

\date{\today}

\begin{abstract}
	The  iterated Crank-Nicolson  is a  predictor-corrector algorithm
	commonly used  in numerical relativity  for the solution  of both
	hyperbolic and parabolic  partial differential equations. We here
	extend  the recent  work  on  the stability  of  this scheme  for
	hyperbolic  equations by  investigating the  properties  when the
	average between  the predicted and corrected values  is made with
	unequal weights  and when  the scheme is  applied to  a parabolic
	equation. We  also propose a variant  of the scheme  in which the
	coefficients in the averages  are swapped between two corrections
	leading to  systematically larger amplification factors  and to a
	smaller numerical dispersion.
\end{abstract}


\maketitle

\section{Introduction}
\label{intro}

	In a recent paper~\cite{teuk_icn_00}, the stability of the
iterated Crank-Nicolson (ICN) method was investigated for the solution of
hyperbolic partial differential equations.  We here extend the work in
ref.~\cite{teuk_icn_00} in three different ways. Firstly, we investigate
the stability properties of the ICN method when the average between the
predicted and corrected values is made with unequal weights as recently
used in refs.~\cite{duez_et_al_03,duez_et_al_04}. Secondly, we apply the
above analysis to a prototypical parabolic partial differential equation,
whose solution is also becoming important within numerical relativity
simulations~\cite{duez_et_al_04}. Finally, we propose a variant of the
scheme, valid for both hyperbolic and parabolic equations, in which the
coefficients in the averages are swapped between two corrections, leading
to larger amplification factors and smaller numerical dispersion.

	The paper is organized as follows: in Sections~\ref{ICN}
and~\ref{sec:gen_icn} we recall the definition of the ICN as a
predictor-corrector method and as a $\theta$-method, respectively. In
Sections~\ref{hypeqs} and~\ref{pareqs}, on the other hand, we discuss the
stability properties of the $\theta$-ICN in the case of hyperbolic and
parabolic equations, respectively. The analysis of the truncation error,
numerical dissipation and dispersion is presented in
Section~\ref{te_nd_d}, while the conclusions are collected in
Section~\ref{conclusions}.

\section{ICN as a predictor-corrector method}
\label{ICN}

	Restricting our discussion to one spatial dimension, hereafter we
will consider a first-order in time partial differential equation of the
type
\begin{equation}
\label{eq:general}
\frac{\partial u(x,t)}{\partial t} = {\cal L}(u(x,t))\;,
\end{equation}
where ${\cal L}$ is a generic quasi-linear partial differential operator
which we assume to contain either first-order or second-order spatial
partial derivatives. Most equations in numerical relativity can be recast
in this form and more complex operators follow from these two cases
(see~\cite{chh05} and references therein).

	After introducing a discretization $\Delta x$ in space and
$\Delta t$ in time, and truncating at the first order the
finite-difference representation of the time derivative in
eq.~(\ref{eq:general})
\begin{equation}
\label{eq:diff1D_FTCS}
\left.\frac{\partial u}{\partial t}\right|_{n,j} =
	\frac{u^{n+1}_{j}-u^{n}_{j}}{\Delta t} + {\cal O}(\Delta t)\;,
\end{equation}
the generic solution of (\ref{eq:general}) can be expressed as
\begin{equation}
u_j^{n+1} = u_j^n + 
	\Delta t \;{\boldsymbol L}\left(u^{n}_{k}\right) \  , 
\end{equation}
where, as usual, $u^{n}_{j} \equiv u(j \Delta x, n \Delta t)$ with $j$
and $n$ integers, and ${\boldsymbol L}$ is the finite-difference form of
the differential operator ${\cal L}$. The spatial index $k$ varies
according to the order at which the operator is represented, with $k=j
\pm 1$ for a second-order accurate, first-order spatial derivative [{\it
cf.} eq.(\ref{fo_spatl})], or with $k=j, j \pm 1$ for a second-order
accurate, second-order spatial derivative [{\it cf.}
eq.(\ref{so_spatl})].

	The ICN scheme discussed in~\cite{teuk_icn_00} is then the
modification of the implicit Crank-Nicolson scheme~\cite{cn_47} as
obtained by truncating, at some point, the following infinite sequence of
predictions and corrections
\begin{subequations}
\label{eq:diff1D_icn_pred}
\begin{gather}
 ^{(1)}\tilde u_j^{n+1} = u_j^n + 
	\Delta t \;{\boldsymbol L}\left(u^{n}_{k}\right)~,               \\
 ^{(1)}\bar u ^{n+1/2}_j \equiv \frac{1}{2} 
	\left( ^{(1)}\tilde u^{n+1}_j + u^n_j \right)~,                   \\
 ^{(2)}\tilde u_j^{n+1} = u_j^n + 
	\Delta t~{\boldsymbol L}\left(^{(1)}\bar u ^{n+1/2}_{k}\right)~,  \\
 ^{(2)}\bar u ^{n+1/2}_j \equiv \frac{1}{2} 
	\left( ^{(2)}\tilde u^{n+1}_j + u^n_j \right)~,                   \\
 ^{(3)}\tilde u_j^{n+1} = u_j^n + 
	\Delta t~{\boldsymbol L}\left(^{(2)}\bar u ^{n+1/2}_{k}\right)~,  \\ 
\vdots \nonumber
\end{gather}
\end{subequations}
where $^{(M)}\bar u_j^{n+1/2}$ is the $M$-th average and $^{(M)}\tilde
u_j^{n+1}$, $^{(M+1)}\tilde u_j^{n+1}$ the $M$-th predicted and corrected
solutions, respectively.

\section{ICN as a ${\boldsymbol \theta}$-method}
\label{sec:gen_icn}

	In the ICN method the $M$-th average is made weighting equally
the newly predicted solution $^{(M)}\tilde u_j^{n+1}$ and the solution at
the ``old'' timelevel'' $u^{n}$. This, however, can be seen as the
special case of a more generic averaging of the type
\begin{equation}\label{eq:gen_ICN}
{^{(M)}\bar u^{n+1/2}} = \theta ~{^{(M)}\tilde u^{n+1}} +
	(1-\theta) u^n \;,
\end{equation}
where $0 < \theta < 1$ is a constant coefficient. Predictor-corrector
schemes using this type of averaging are part of a large class of
algorithms named \emph{$\theta$-methods}~\cite{richt_mort_67}, and we
refer to the ICN generalized in this way as to the \mbox{``$\theta$-ICN''
method}.

	A different and novel generalization of the $\theta$-ICN can be
obtained by {\it swapping} the averages between two subsequent corrector
steps, so that in the $M$-th corrector step
\begin{equation}
\label{eq:swap_icn_1stcorr}
{^{(M)}\bar u^{n+1/2}} = (1 - \theta)\,{^{(M)}\tilde u^{n+1}} + \theta
	u^n~,
\end{equation}
while in the $(M+1)$-th corrector step
\begin{equation}
\label{eq:swap_icn_2ndcorr}
{^{(M+1)}\bar u^{n+1/2}} = \theta\,{^{(M+1)}\tilde u^{n+1}} + 
	(1-\theta)u^n~.
\end{equation}
Note that as long as the number of iterations is even, the sequence in
which the averages are computed is irrelevant. Indeed, the weights
$\theta$ and $1-\theta$ in
eqs.~\eqref{eq:swap_icn_1stcorr}--\eqref{eq:swap_icn_2ndcorr} could be
inverted and all of the relations discussed hereafter for the swapped
weighted averages would continue to hold after the transformation $\theta
\to 1 - \theta$.
	
	Although the properties of the $\theta$-ICN do not seem to have
been discussed before, the scheme has already found application in
numerical relativity calculations, where it has been used with a
coefficient $\theta = 0.6$ in the solution of the relativistic
hydrodynamics equations for ideal~\cite{duez_et_al_03} and viscous
fluids~\cite{duez_et_al_04}. In these works it was found that the use of
a weighting coefficient different from 1/2 yielded ``an improved
stability''. In Sect.~\ref{te_nd_d} we will show that such a choice has
effectively only increased the numerical dissipation of the scheme.

\section{Hyperbolic equations}
\label{hypeqs}

	To discuss the properties of the $\theta$-ICN we consider as
model hyperbolic equation the one-dimensional advection equation
\begin{equation}
\label{eq:adv1D}
\frac{\partial u}{\partial t} + v \frac{\partial u}{\partial x} = 0~,
\end{equation}
where $v$ is a constant coefficient. A second-order accurate
finite-difference representation of the right-hand-side of
eq.~\eqref{eq:adv1D} is then easy to derive and has the form
\begin{equation}
\label{fo_spatl}
{\boldsymbol L}(u^n_{j \pm 1}) = \frac{u^{n}_{j+1} - u^{n}_{j-1}}
	{2 \Delta x} + {\cal O}(\Delta x^2)~.
\end{equation}

\subsection{Constant Arithmetic Averages}

	Using a von Neumann stability analysis, Teukolsky has shown that
for a hyperbolic equation the ICN scheme with $M$ iterations has an
amplification factor~\cite{teuk_icn_00}
\begin{equation}
\label{eq:xi_hyp_icn}
^{(M)}\xi = 1 + 2 \sum_{n=1}^{M} 
	\left( - {\rm i} \beta \right)^{n}\;,
\end{equation}
where $\beta \equiv v [\Delta t / (2\Delta x)] \sin k \Delta
x$~\footnote{Note that we define $\beta$ to have the opposite sign of the
corresponding quantity defined in ref.~\cite{teuk_icn_00}}. We recall
that in a von Neumann stability analysis the eigenmodes of the
finite-difference equations are expressed as $u^n_j = \xi^n e^{{\rm i} k
j \Delta x}$, where $k$ is a real spatial wavenumber and $\xi=\xi(k)$ is
a complex number. Stability then requires that $| \xi | \equiv \sqrt{\xi
\xi^*} \leq 1$ and in eq.~\eqref{eq:xi_hyp_icn} this leads to an
alternating pattern in the number of iterations. More specifically, zero
and one iterations yield an unconditionally unstable scheme, while two
and three iterations a stable one provided that $\beta ^2 \leq 1$; four
and five iterations lead again to an unstable scheme and so
on. Furthermore, because the scheme is second-order accurate from the
first iteration on, Teukolsky's suggestion when using the ICN method for
hyperbolic equations was that two iterations should be used {\it and no
more}~\cite{teuk_icn_00}. This is the number of iterations we will
consider hereafter.

\subsection{Constant Weighted Averages}

\begin{figure*}[btp]
\centering 
\includegraphics[width = 0.495\textwidth]{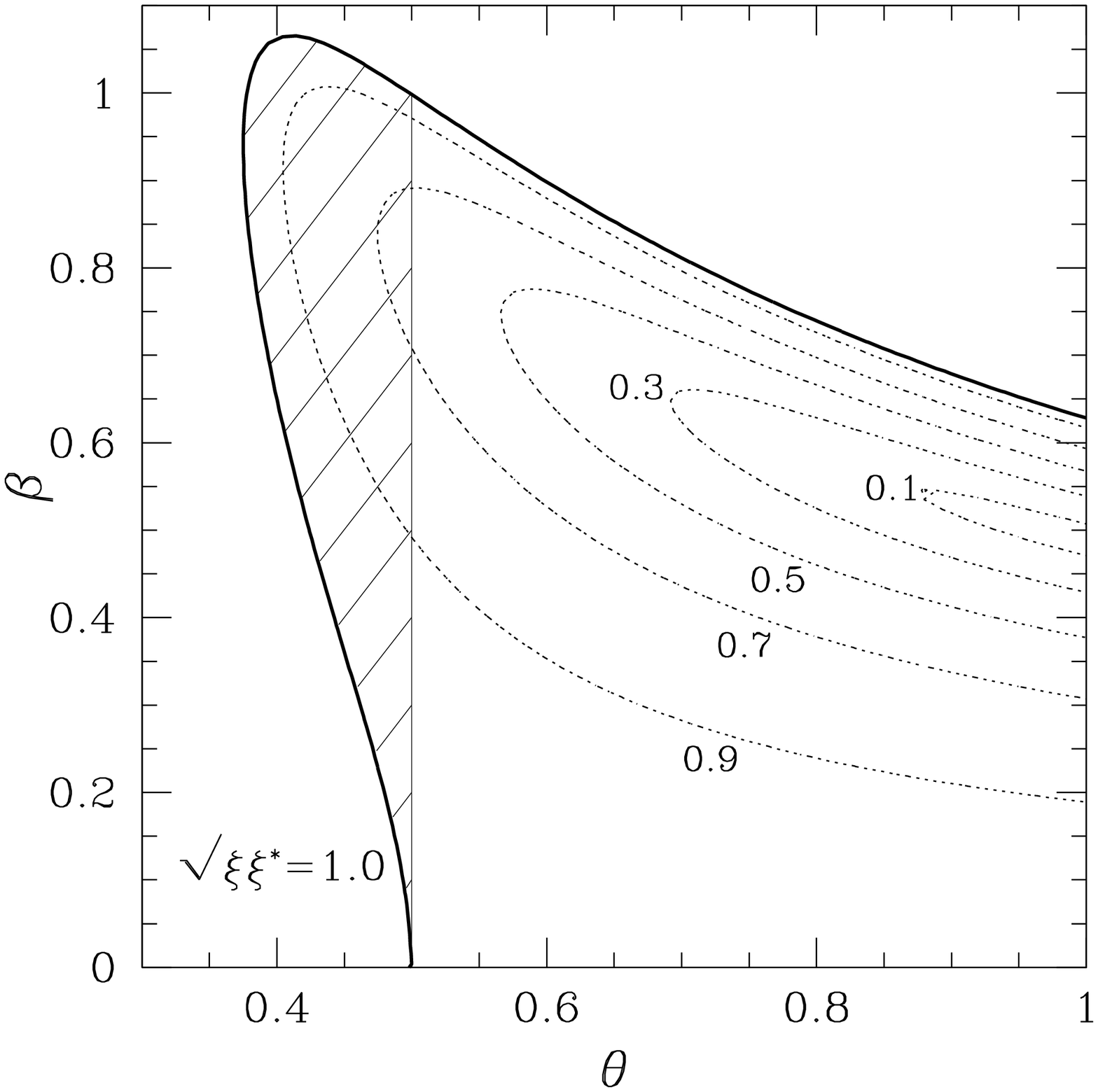}
\includegraphics[width = 0.495\textwidth]{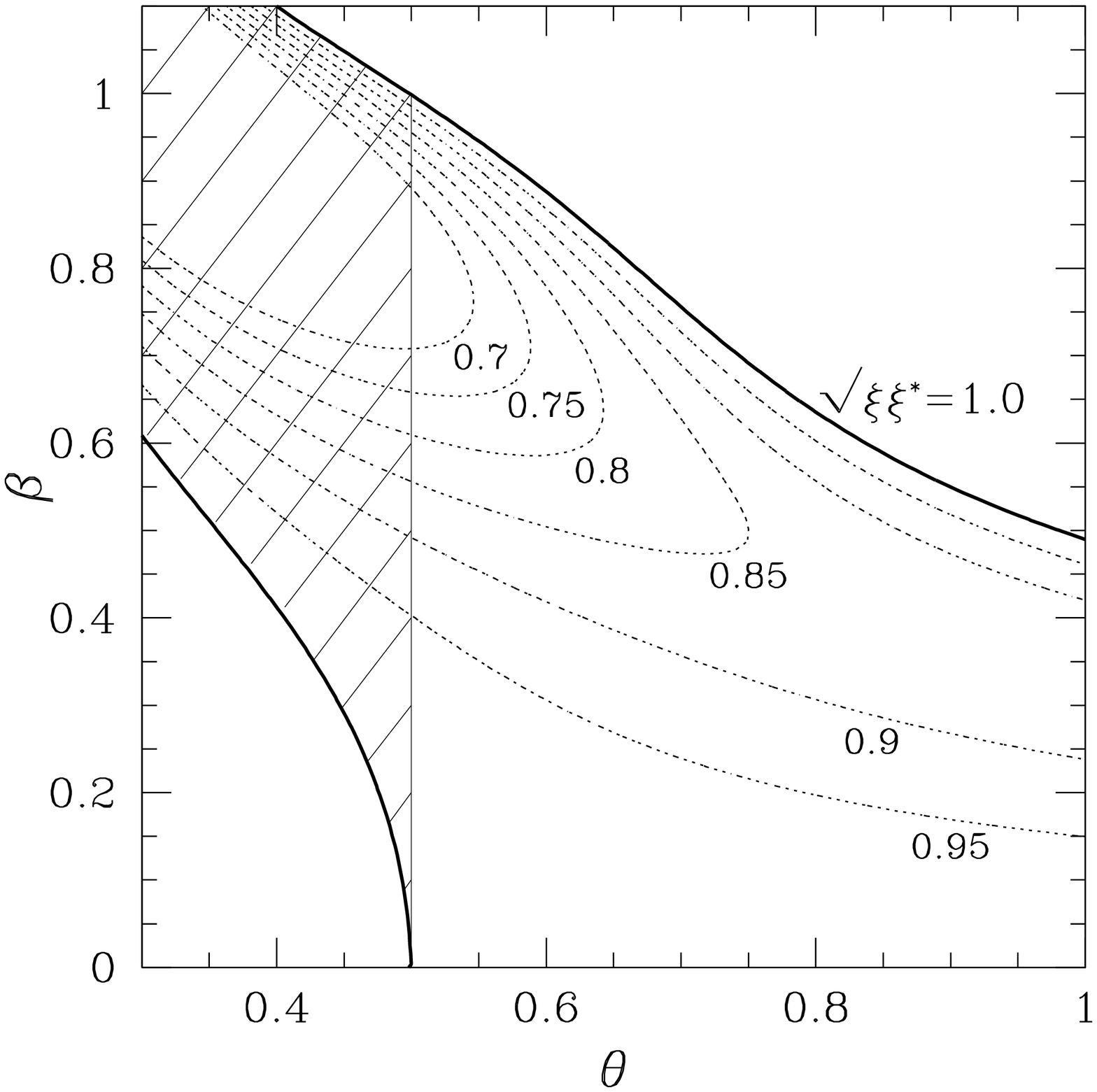}
\caption{{\it Left panel:} stability region in the ($\theta,\beta$) plane
for the two-iterations $\theta$-ICN for the advection
equation~\eqref{eq:adv1D}. Thick solid lines mark the limit at which
$|\xi|=1$, while the dotted contours indicate the values of the
amplification factor in the stable region. The shaded area for $\theta <
1/2$ refers to solutions that are linearly
unstable~\cite{bar-gri-hig_00}. {\it Right panel:} same as in the left
panel but when the averages between two corrections are swapped. Note
that the amplification factor in this case is less sensitive on $\theta$
and always larger than the corresponding amplification factor in the left
panel.}
\label{fig:allowed_alpha}
\end{figure*}

	Performing the same stability analysis for a $\theta$-ICN is only
slightly more complicated and truncating at two iterations the
amplification factor is found to be
\begin{equation}
\label{eq:gen_icn_hyp_xi2}
\xi = 1 - 2{\rm i} \beta - 4 \beta^2 \theta +
	8 {\rm i} \beta^3 \theta^2 \ ~, 
\end{equation}
where $\xi$ is a shorthand for ${^{(2)}\xi}$. The stability condition in
this case translates into requiring that
\begin{equation}
\label{eq:gen_icn_hyp_stab}
16 \beta^4 \theta^4 - 4 \beta^2\theta^2 - 2\theta + 1 \leq 0\;,
\end{equation}
or, equivalently, that for $\theta > 3/8$ 
\begin{equation}
\label{eq:hyp_icn_allowed_alpha}
\frac{\sqrt{\frac{1}{2} - \sqrt{2\theta - \frac{3}{4}}}}{2 \theta} 
	\leq \beta
	\leq \frac{\sqrt{\frac{1}{2} + \sqrt{2\theta - 
	\frac{3}{4}}}}{2\theta}\;,
\end{equation}
which reduces to $\beta^2 \leq 1$ when $\theta = 1/2$. Because the
condition \eqref{eq:hyp_icn_allowed_alpha} must hold for every wavenumber
$k$, we consider hereafter $\beta \equiv v \Delta t / (2\Delta x)$ and
show in the left panel of Fig.~\ref{fig:allowed_alpha} the region of
stability in the ($\theta,\beta$) plane. The thick solid lines mark the
limit at which $|\xi|= 1$, while the dotted contours indicate the
different values of the amplification factor in the stable region.

	A number of comments are worth making. Firstly, although the
condition~\eqref{eq:hyp_icn_allowed_alpha} allows for weighting
coefficients $\theta < 1/2$, the $\theta$-ICN is stable only if $\theta
\geq 1/2$. This is a known property of the weighted Crank-Nicolson
scheme~\cite{richt_mort_67} and inherited by the $\theta$-ICN. In
essence, when $\theta \ne 1/2$ spurious solutions appear in the
method~\cite{stu-pep_91} and these solutions are linearly unstable if
$\theta < 1/2$, while they are stable for $\theta >
1/2$~\cite{bar-gri-hig_00} (An alternative and simpler explanation is
also presented in Sect.~\ref{te_nd_d}). For this reason we have shaded
the area with $\theta < 1/2$ in the left panel of
Fig.~\ref{fig:allowed_alpha} to exclude it from the stability
region. Secondly, the use of a weighting coefficient $\theta > 1/2$ will
still lead to a stable scheme provided that the timestep ({\it i.e.,}
$\beta$) is suitably decreased. Finally, as the contour lines in the left
panel of Fig.~\ref{fig:allowed_alpha} clearly show, the amplification
factor can be very sensitive on $\theta$.

\subsection{Swapped weighted averages}
\label{sec:swap_icn}

	The calculation of the stability of the $\theta$-ICN when the
weighted averages are swapped as in eqs.~\eqref{eq:swap_icn_1stcorr}
and~\eqref{eq:swap_icn_2ndcorr} is somewhat more involved; after some
lengthy but straightforward algebra we find the amplification factor to
be
\begin{eqnarray}
\label{eq:swap_icn_hyp_xi2}
\xi &=& 1 - 2 {\rm i} \beta  - 4 \beta^2 \theta +
	8 {\rm i} \beta^3 \theta (1-\theta)\;,
\end{eqnarray}
which differs from~\eqref{eq:gen_icn_hyp_xi2} only in that the $\theta^2$
coefficient of the ${\cal O}(\beta^3)$ term is replaced by
$\theta(1-\theta)$. The stability requirement $|\xi| \leq 1$ is now
expressed as
\begin{equation}
\label{eq:swap_icn_hyp_stab}
16 \beta^4 \theta^2 (1- \theta)^2 - 4 \beta^2 \theta(2- 3\theta) -
	2\theta + 1 \leq 0\;.
\end{equation}
Solving the condition~\eqref{eq:swap_icn_hyp_stab} with respect to
$\beta$ amounts then to requiring that 
\begin{subequations}
\begin{gather}
\label{eq:swap_icn_hyp_stab_bis}
\beta \geq  \frac{\sqrt{2 - 3\theta - \sqrt{4\theta - 11\theta^2 +
	8\theta^3}}}{2(1-\theta)\sqrt{2\theta}}\;,
\\
\beta \leq  \frac{\sqrt{2 - 3\theta + \sqrt{4\theta - 11\theta^2 +
	8\theta^3}}}{2(1-\theta)\sqrt{2\theta}}\;,
\end{gather}
\end{subequations}
which is again equivalent to $\beta^2 \leq 1$ when $\theta = 1/2$. The
corresponding region of stability is shown in right panel of
Fig.~\ref{fig:allowed_alpha} and should be compared with left panel of
the same Figure. Note that the average-swapping has now considerably
increased the amplification factor, which is always larger than the
corresponding one for the $\theta$-ICN in the relevant region of
stability ({\it i.e.,} for $1/2 \leq \theta \leq 1$~\footnote{Of course,
when the order of the swapped averages is inverted from the one shown in
eqs.~\eqref{eq:swap_icn_1stcorr}--\eqref{eq:swap_icn_2ndcorr} the
stability region will change into $0 \leq \theta \leq 1/2$.}).
	
\section{Parabolic equations}
\label{pareqs}

	We next extend the stability analysis of the $\theta$-ICN to the
a parabolic partial differential equation and use as model equation the
one-dimensional diffusion equation
\begin{equation}
\label{eq:diff1D}
\frac{\partial u}{\partial t} - D\frac{\partial^2 u}{\partial x^2} = 0\;,
\end{equation}
where $D$ is a constant coefficient which must be positive for the
equation to be well-posed.

	Parabolic equations are commonly solved using implicit methods
such as the Crank-Nicolson, which is unconditionally stable and thus
removes the constraints on the timestep [{\it i.e.,} $\Delta t \approx
{\cal O}(\Delta x ^2)$] imposed by explicit
schemes~\cite{press_et_al_97}. In multidimensional calculations, however,
or when the set of equations is of mixed hyperbolic-parabolic type,
implicit schemes can be cumbersome to implement since the resulting
system of algebraic equations does no longer have simple and tridiagonal
matrices of coefficients. In this case, the most conveniente choice may
be to use an explicit method such as the ICN.

	Also in this case, the first step in our analysis is the
derivation of a finite-difference representation of the right-hand-side
of eq.~\eqref{eq:diff1D} which, at second-order, has the form
\begin{equation}
\label{so_spatl}
{\boldsymbol L}(u^n_{j,j \pm 1}) = \frac{u^{n}_{j+1} - 2 u^{n}_{j}
  	+u^{n}_{j-1}}{\Delta x^2} + {\cal O}(\Delta x^2)\;.
\end{equation}

\subsection{Constant Arithmetic Averages}
\label{cav_p}

	Next, we consider first the case with constant arithmetic
averages ({\it i.e.,} $\theta = 1/2$) and the expression for the
amplification factor after $M$-iterations is then purely real and given
by
\begin{equation}
\label{eq:xi_diff_icn}
^{(M)}\xi = 1 + 2 \sum_{n=1}^{M} \left( - \gamma \right)^{n}\;,
\end{equation}
where $\gamma \equiv (2 D \Delta t /\Delta x^2) \sin^2 (k \Delta
x/2)$. Requiring now for stability that $\sqrt{\xi^2} \leq 1$ and bearing
in mind that
\begin{equation}
-1 \leq \sum_{n=0}^{M} \left( - \gamma \right)^{n+1} \leq 0\;, 
	\hskip 0.5cm {\rm for\ \ } \gamma \leq 1 \;, 
\end{equation}
we find that the scheme is stable for \emph{any} number of iterations
provided that $\gamma \leq 1$. Furthermore, because the scheme is
second-order accurate from the first iteration on, our suggestion when
using the ICN method for parabolic equations is that one iteration should
be used {\it and no more}. In this case, in particular, the ICN method
coincides with a FTCS scheme~\cite{press_et_al_97}.

\begin{figure*}[bpt]
\centering
\includegraphics[width = 0.495\textwidth]{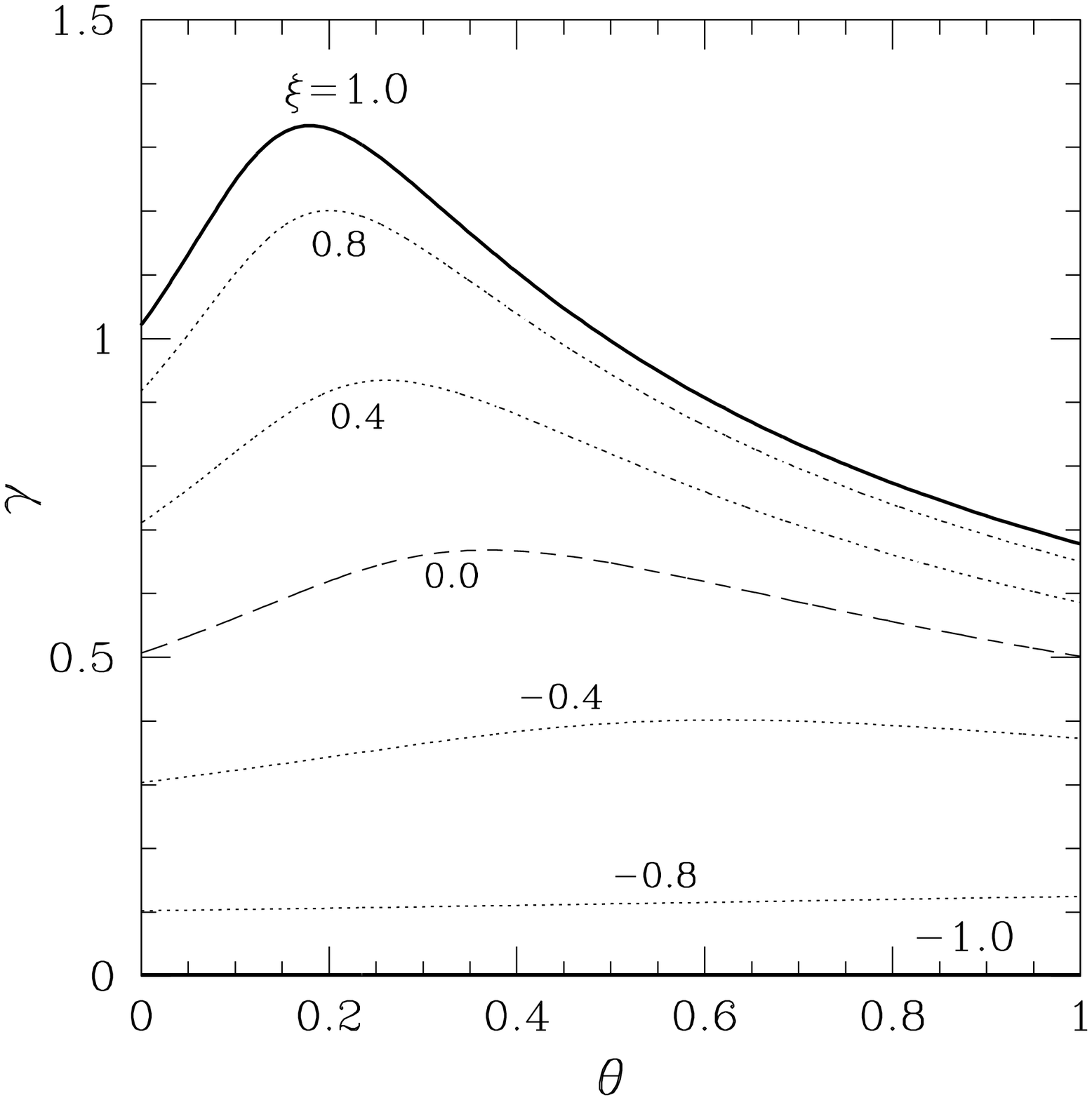}
\includegraphics[width = 0.495\textwidth]{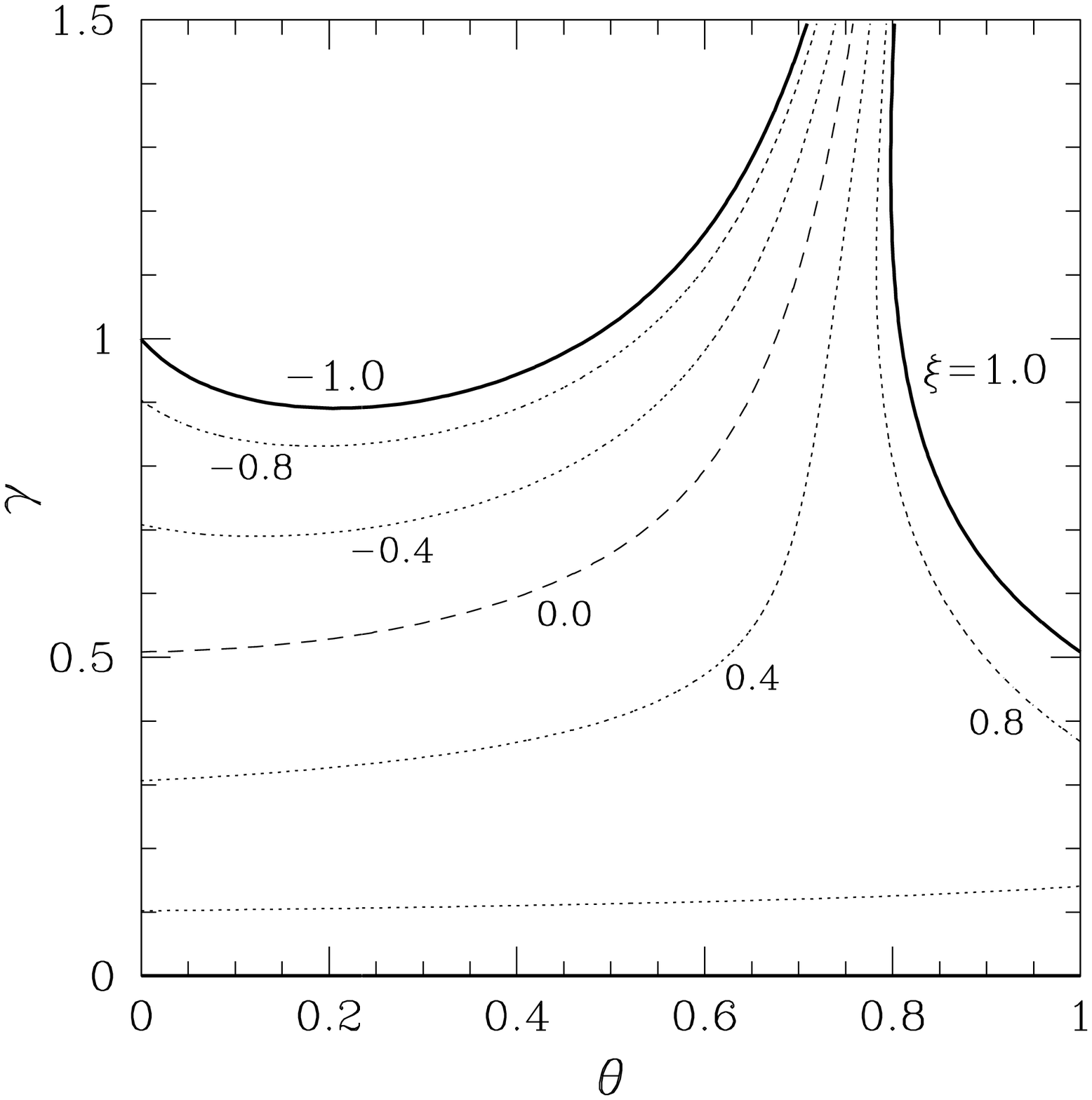}
\caption{{\it Left panel:} stability region in the ($\theta,\gamma$)
plane for the two-iterations $\theta$-ICN for the diffusion
equation~\eqref{eq:diff1D}. Thick solid lines mark the limit at which
$\xi^2=1$, while the dotted contours indicate the values of the
amplification factor in the stable region. {\it Right panel:} same as in
the left panel but with swapping the averages between two corrections.}
\label{fig:allowed_gamma}
\end{figure*}

	Note that the stability condition \hbox{$\gamma \leq 1$}
introduces again a constraint on the timestep that must be $\Delta t \leq
\Delta x^2/(2D)$ and thus ${\cal O}(\Delta x^2)$. As a result and at
least in this respect, the ICN method does not seem to offer any
advantage over other explicit methods for the solution of a parabolic
equation~\footnote{Note that also the Dufort-Frankel
method~\cite{duf_53}, usually described as unconditionally stable, does
not escape the timestep constraint $\Delta t \approx {\cal O}(\Delta x
^2)$ when a consistent second-order accurate solution is
needed~\cite{smith_85}.}.

\subsection{Constant Weighted Averages}

	We next consider the stability of the $\theta$-ICN method but
focus our attention on a two-iterations scheme since this is the number
of iterations needed in the solution of the parabolic part in a mixed
hyperbolic-parabolic equation when, for instance, operator-splitting
techniques are adopted~\cite{press_et_al_97}. In this case, the
amplification factor is again purely real and given by
\begin{eqnarray}
\label{eq:gen_icn_diff_xi2}
\xi &=& 1 - 2 \gamma + 4 \gamma^2 \theta - 8 \gamma^3 \theta^2\;,
\end{eqnarray}
so that stability is achieved if
\begin{equation}
\label{eq:gen_icn_diff_stab_syst}
0 \leq \gamma\left(1 - 2 \theta \gamma + 4 \theta^2 \gamma^2 \right)\leq
	1 ~.
\end{equation}
Since $\gamma > 0$ by definition, the left inequality is always
satisfied, while the right one is true provided that, for $\gamma < 4/3$,
\begin{equation}
\label{eq:diff_icn_allowed_gamma}
\frac{\gamma - \sqrt{\gamma(4-3\gamma)}}{4\gamma^2} \leq \theta \leq
\frac{\gamma + \sqrt{\gamma(4-3\gamma)}}{4\gamma^2}\;.
\end{equation}

	The stability region described by the condition
\eqref{eq:diff_icn_allowed_gamma} is shown in the left panel of
Fig.~\ref{fig:allowed_gamma} for $\sin k \Delta x = 1$ and illustrates
that the scheme is stable for any value $0 \leq \theta \leq 1$, and also
that slightly larger timesteps can be taken when $\theta \simeq 0.2$.

\subsection{Swapped Weighted Averages}
\label{subsec:swap_icn_par}

	After some lengthy algebra the calculation of the amplification
factor for the $\theta$-ICN method with swapped weighted averages yields
\begin{equation}
\label{eq:swap_icn_diff_xi2}
\xi = 1 - 2 \gamma + 4 \gamma^2 \theta - 
	8 \gamma^3 \theta (1-\theta)\;,
\end{equation}
and stability is then given by
\begin{equation}
\label{eq:swap_icn_diff_stab}
-1 \leq 1 - 2 \gamma + 4 \gamma^2 \theta - 
	8 \gamma^3 \theta (1-\theta) \leq 1\;.
\end{equation}
Note that none of the two inequalities is always true and in order to
obtain analytical expressions for the stable region we solve the
condition \eqref{eq:swap_icn_diff_stab} with respect to $\theta$ and
obtain
\begin{subequations}
\label{eq:swap_icn_allowed_beta}
\begin{gather}
\theta \leq \frac{2\gamma - 1 + \sqrt{4 \gamma^2 - 4 \gamma + 5}}{4
	\gamma}~, \label{eq:swap_icn_allowed_beta_a} 
\\ 
\label{eq:swap_icn_allowed_beta_c}
\theta \leq \frac{\gamma (2\gamma - 1) - \sqrt{\gamma \left(
	4\gamma^3 - 4\gamma^2 +5\gamma - 4\right)}} {4\gamma^2}\;,
\\ 
\label{eq:swap_icn_allowed_beta_d}
\theta \geq \frac{\gamma (2\gamma - 1) + \sqrt{\gamma \left(
	4\gamma^3 - 4\gamma^2 +5\gamma - 4\right)}} {4\gamma^2}\;.
\end{gather}
\end{subequations}
The resulting stable region for $\sin k \Delta x = 1$ is plotted in the
right panel of Fig.~\ref{fig:allowed_gamma} and seems to suggest that
arbitrarily large values of $\gamma$ could be considered when $\theta
\gtrsim 0.6$ It should be noted, however, that the amplification factor
is also severely reduced as larger values of $\gamma$ are used and indeed
it is essentially zero in the limit $\theta \to 1$. 

\section{Truncation error, dissipation and dispersion}
\label{te_nd_d}

	Although not often appreciated, the $\theta$-ICN method is only
first-order accurate in time as an obvious consequence of the first-order
approximation in the time derivative [{\it cf.}
eq.~\eqref{eq:diff1D_FTCS}]. However, this is not true if $\theta =1/2$,
in which case the method becomes second-order in both space and time.

	To appreciate this in the case of the advection
equation~\eqref{eq:adv1D}, we report the finite-difference expressions
for the time and spatial derivatives in eq.~\eqref{eq:adv1D}, writing out
explicitly the coefficients of the ${\cal O}(\Delta t)$ and ${\cal
O}(\Delta x^2)$ terms
\begin{eqnarray}
\label{eq:u_t-fd-bis}&&\frac{u^{n+1}_j - u^n_j}{\Delta t} = 
	\left . \frac{\partial u}{\partial t} \right|_{n,j} \!\!\!+
	\left. \frac{1}{2} \frac{\partial^2 u}{\partial t^2}\right|_{n,j} 
	\!\!\!\!\!\!\Delta t+ 
	{\cal O}\left( \Delta t^2 \right) \;,
\\ \nonumber \\
&&\frac{u^n_{j+1} - u^n_{j-1}}{2 \Delta x} = \left. \frac{\partial
	u}{\partial x} \right|_{n,j} \!\!\!+
	\left. \frac{1}{6} \frac{\partial^3 u}{\partial x^3} 
	\right|_{n,j} \!\!\!\!\!\!\Delta x^2 +
	{\cal O}(\Delta x ^4)\;.
\nonumber \\ 
\end{eqnarray}
The resulting local truncation error is then
\begin{eqnarray}
\label{eq:adv_icn_loc.t.er}
e_{_{\rm T}} &=& \left( \frac{1}{2} - \theta \right) v 
	\left.\frac{\partial^2 u}{\partial x \partial t} 
	\right|_{n,j} \!\!\!\!\!\!\Delta t - 
	\frac{v}{6} \left.\frac{\partial^3 u}{\partial x^3} 
	\right|_{n,j} \!\!\!\!\!\!\Delta x^2  
\nonumber \\
&& 
	- v^3 \theta^2 \left.\frac{\partial^3 u}{\partial x^3} 
	\right|_{n,j} \!\!\!\!\!\!\Delta t^2  
	- \frac{1}{6}\left.\frac{\partial^3 u}{\partial t^3} 
	\right|_{n,j} \!\!\!\!\!\! \Delta t^2  
\nonumber \\
&& 
	+ \ {\cal O}\left( \Delta t^3, \Delta x^3 \right)~,
\end{eqnarray}
clearly indicating that the $\theta$-ICN is generally only first-order
accurate in time, becoming second-order if \mbox{$\theta = 1/2$}. The
truncation error is also useful to quantify the numerical dissipation and
dispersion inherent to the $\theta$-ICN method. Using
eq.~\eqref{eq:adv1D} to replace the time derivative with a spatial one,
in fact, eq.~\eqref{eq:adv_icn_loc.t.er} shows that the $\theta$-ICN
introduces a dissipative term proportional to $\partial^2 u/\partial x^2$
and with coefficient
\begin{equation}
\label{eq:dissipation_adv}
\epsilon_{\rm adv} = \left( \theta - \frac{1}{2} \right) v^2 
	\Delta t~.
\end{equation}
In other words, the $\theta$-ICN is intrinsically dissipative, with a
dissipation coefficient that is generically ${\cal O}(\Delta t)$ and
${\cal O}(\Delta t^3, \Delta x^3)$ only when $\theta = 1/2$. Furthermore,
it is now apparent why $\theta$ must be larger or equal to $1/2$; any
choice different from this, in fact, would change the sign of
$\epsilon_{\rm adv} $, leading to an ill-posed equation with
exponentially growing solutions [{\it cf.}  eq.~\eqref{eq:diff1D}].

	Expression~\eqref{eq:dissipation_adv} also clarifies the
behaviour found in refs.~\cite{duez_et_al_03, duez_et_al_04}. Since
stability in a numerical scheme is either gained or lost but cannot be
``improved'', the use of a weighting coefficient $\theta > 1/2$ (and of a
suitable timestep) has simply the effect of increasing the numerical
dissipation of the scheme. Of course, this is often a desirable feature
to suppress the growth of instabilities, as in the case of the
Lax-Friedrichs scheme, whose numerical dissipation stabilizes the
otherwise unconditionally unstable FTCS scheme~\cite{press_et_al_97}.

	An alternative route to a second-order, moderately dissipative
scheme is to choose
\begin{equation}
\label{eq:theta_accu_adv}
\theta = \frac{1}{2} + \frac{\Delta x^2}{v \Delta t}\;,
\end{equation}
with ${\Delta x^2}/({v \Delta t}) \leq 1/2$, so that the leading
error-term in~\eqref{eq:adv_icn_loc.t.er} becomes again ${\cal O}(\Delta
t^2, \Delta x^2)$. A prescription of the type~\eqref{eq:theta_accu_adv}
may be the {\it optimal} one for the $\theta$-ICN method as it provides a
small amount of numerical dissipation {\it and} reduces the truncation
error.

	The truncation error~\eqref{eq:adv_icn_loc.t.er} also indicates
that the $\theta$-ICN introduces a dispersive term proportional to
$\partial^3 u/\partial x^3$ given by
\begin{equation}
\label{eq:dispersion_adv}
\chi_{\rm adv} = \left( \frac{1}{6} - \theta^2 - 
	\frac{1}{24 \beta^2}\right) v^3 
	\Delta t^2  \;,
\end{equation}
and responsible, for instance, for different propagation speeds of the
Fourier modes in the initial data ({\it i.e.,} phase drifts).

	All what discussed so far for the $\theta$-ICN scheme continues
to hold also when the averages are swapped, the only difference being
that the dispersive contribution is instead given by 
\begin{equation}
\label{eq:dispersion_adv_swap}
\chi_{\rm adv} = \left( \frac{1}{6} - \theta + \theta^2 - 
	\frac{1}{24 \beta^2}\right) v^3
	\Delta t^2 \;,
\end{equation}
and is therefore smaller for $\theta > 1/2$, making this variant to the
$\theta$-ICN preferable overall.

	Similar calculations can be carried out also for the parabolic
equation~\eqref{eq:diff1D} and the local truncation error in this case is
\begin{equation}
\label{eq:diff_icn_loc.t.e}
e_{_{\rm T}} = \left( \theta - \frac{1}{2} \right)D 
	\left. \frac{\partial^3 u}{\partial t \partial x^2} 
	\right|_{n,j} \!\!\!\!\!\! \Delta t + 
	{\cal O} \left( \Delta t^2, \Delta x^2 \right)\;,
\end{equation}
indicating that mathematically the $\theta$-ICN is again only first-order
accurate in time, with second-order accuracy being recovered for $\theta
= 1/2$. However, stability requires that \mbox{$\Delta t = {\cal O}
\left( \Delta x^2 \right)$} ({\it cf.}  Sect.~\ref{cav_p}) and thus the
truncation error is effectively $e_{_{\rm T}} = {\cal O} \left( \Delta
x^2 \right)$ for all of the allowed values $0 \leq \theta \leq 1$.

	Finally, using eq.~\eqref{eq:diff1D} to the replace the time
derivative in~\eqref{eq:diff_icn_loc.t.e} shows that the $\theta$-ICN for
a parabolic equation has an additional dissipative term proportional to
$\partial^4 u/\partial x^4$ with coefficient
\begin{equation}
\label{eq:dispersion_diff_swap}
\epsilon_{\rm diff} = \left( \theta - \frac{1}{2} \right)D^2 
	\Delta t \;,
\end{equation}
which is again zero only for $\theta = 1/2$.

\begin{figure}[bpt]
\centering 
\includegraphics[width = 0.48\textwidth]{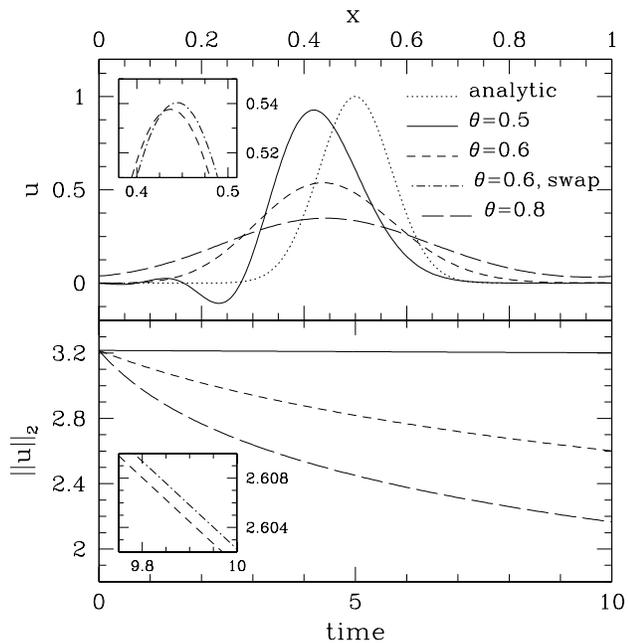}

\caption{{\it Upper panel:} Solution of the advection equation
  \eqref{eq:adv1D} using the $\theta$-ICN method and shown after 10
  crossing times. Different curves refer to either the analytic solution
  or to the numerical ones with different weighting coefficients. The
  small inset, instead, shows the smaller diffusion and dispersion
  obtained when the averages are swapped (see main text for
  details). {\it Lower panel:} L$_2$ norms of the solutions in the upper
  panel plotted as a function of time.}
\label{fig:example_adv}
\end{figure}

	As a purely representative example we show in
Fig.~\ref{fig:example_adv} the application of the $\theta$-ICN method for
the solution of the advection equation \eqref{eq:adv1D} with $v=1$ and
\mbox{$\beta = 0.6$} ({\it cf.}  Fig.~\ref{fig:allowed_alpha}). The
numerical domain has length $1.0$ and was covered with 200 equally spaced
gridpoints. The initial solution, given by a Gaussian centred at $x=0.5$
and with variance $0.1$, was evolved for $10$ crossing times using
periodic boundary conditions. Different curves in the upper panel refer
to either the analytic solution at the final time (dotted line) or to the
numerical solutions as obtained with different weighting
coefficients. Note that already with $\theta=1/2$ (solid line) the
numerical solution is slightly diffused but suffers from considerable
dispersion as apparent from the considerable ``delay'' and the presence
of negative values to the left of the maximum. These dispersion errors
can be reduced if larger values of the weighting coefficients are used as
indicated by the short-dashed and long-dashed lines referring to
$\theta=0.6$ and $\theta=0.8$, respectively. This improvement, however,
also comes with a larger dissipation and truncation error (as mentioned
in Sect.~\ref{te_nd_d}, the system is just first-order in time with
$\theta > 1/2$)~\footnote{Note that for all values of $\theta \geq 1/2$ a
smaller dispersion can be obtained for smaller values of $\beta$ and
hence of $\Delta t$; {\it cf.}  eqs.~\eqref{eq:dispersion_adv} and
\eqref{eq:dispersion_adv_swap}}. This is particularly evident when
considering the evolution of the L$_2$ norms of the solutions as reported
in the lower panel of Fig.~\ref{fig:example_adv}. It is interesting to
note that for $\theta=0.8$ the L$_2$ norm of the solution has been
reduced of about 25\% after 10 crossing times, while this decrease is
less than 1\% when $\theta=1/2$.

	Finally, the two small insets in Fig.~\ref{fig:example_adv} offer
a comparison in the solutions for $\theta=0.6$ when the coefficients in
the averages are either held constant (short-dashed lines) or swapped
between two subsequent corrector steps (dot-dashed lines). Although the
difference is rather small for the selected set of parameters, it is
evident that the swapping of the coefficients has the effect of
decreasing both the dispersion (the dot-dashed line in the upper inset
has a smaller ``delay'') and the diffusion (at any given time the
dot-dashed line in the lower inset has a larger value).

\section{Conclusions}
\label{conclusions}

	We have extended the recent work on the properties of the ICN
scheme for hyperbolic equations by investigating the stability properties
when it is treated as a $\theta$-method, {\it i.e.,} when the average
between the predicted and corrected values is made with unequal weights.
In addition we have studied the properties of the $\theta$-ICN method for
a model parabolic equation and proposed a variant of the scheme, valid
for both hyperbolic and parabolic equations, in which the unequal
coefficients coefficients in the averages are swapped between two
subsequent corrector steps. This novel approach leads to amplification
factors that are systematically larger than those found in the
$\theta$-ICN method and to a smaller numerical dispersion.

	Overall, our results indicate that although generally only
first-order accurate in time, the $\theta$-ICN method is a flexible
approach to the time-integration of partial differential equations,
particularly when these are of mixed hyperbolic-parabolic type. Because
the use of unequal coefficients in the average provides a small but
nonzero amount of numerical dissipation, this could prove useful in
numerical relativity calculations which may suffer from the development
of numerical instabilities and for which lower-order evolution schemes
are an acceptable compromise between accuracy and stability.

\section*{ACKNOWLEDGMENTS}
It is a pleasure to thank S. Teukolsky and I. Hawke for useful comments.

\bibliographystyle{apsrev}
\bibliography{biblio}

\end{document}